\documentclass[12pt]{article}

\usepackage{times}
\usepackage{graphicx}

\title{Nanoscale superconducting memory based on the kinetic inductance of asymmetric nanowire loops}

\author
{Andrew Murphy,$^{1}$ Dmitri V. Averin,$^{2}$ Alexey Bezryadin$^{1\ast}$\\
\\
\normalsize{$^{1}$Department of Physics, University of Illinois at Urbana-Champaign,}\\
\normalsize{ Urbana, Illinois 61801, USA,}\\
\normalsize{$^{2}$Department of Physics and Astronomy, Stony Brook University, SUNY,}\\
\normalsize{ Stony Brook, NY 11794-3800, USA}\\
\\
\normalsize{$^\ast$To whom correspondence should be addressed; E-mail:  bezryadi@illinois.edu.}
}

\date{}

\begin{document}

\baselineskip24pt

\maketitle

\begin{abstract}
The demand for low-dissipation nanoscale memory devices is as strong as ever. As Moore's Law is staggering, and the demand for a low-power-consuming supercomputer is high, the goal of making information processing circuits out of superconductors is one of the central goals of modern technology and physics. So far, digital superconducting circuits could not demonstrate their immense potential. One important reason for this is that a dense superconducting memory technology is not yet available. Miniaturization of traditional superconducting quantum interference devices is difficult below a few micrometers because their operation relies on the geometric inductance of the superconducting loop. Magnetic memories do allow nanometer-scale miniaturization, but they are not purely superconducting \cite{spin4}. Our approach is to make nanometer scale memory cells based on the kinetic inductance (and not geometric inductance) of superconducting nanowire loops, which have already shown many fascinating properties \cite{Aprili, Hopkins-Science}. This allows much smaller devices and naturally eliminates magnetic-field cross-talk. We demonstrate that the vorticity, i.e., the winding number of the order parameter, of a closed superconducting loop can be used for realizing a nanoscale nonvolatile memory device. We demonstrate how to alter the vorticity in a controlled fashion by applying calibrated current pulses. A reliable read-out of the memory is also demonstrated. We present arguments that such memory can be developed to operate without energy dissipation.
\end{abstract}

Power management and cooling demands of high performance processors have become one of the main obstacles to further progress of the computing devices. Thus development of the superconductor-based cryogenic computers which appear particularly suitable for overcoming these problems attracts much attention \cite{Manheimer}. Nanoscale low-dissipation memory that could be integrated naturally with superconducting circuits remains one of the most essential elements that still needs to be demonstrated. Typical "single-flux quanta" (SFQ) digital superconductor devices are based on manipulation of individual quanta of magnetic flux in circuits composed of Josephson junctions and inductive loops \cite{rsfq}, and recently made much progress towards the large-scale practical logic circuits -- see, e.g. \cite{Filippov,nsquid,rql,esfq}. However, direct applications of the SFQ principles to memory devices (see, e.g., \cite{mem0,mem1}) remain not quite competitive with other approaches because of the relatively large size, in the micrometer range, of the memory cells, determined by both the size of the Josephson junctions and geometric inductances. This motivates a search for hybrid memory based either on direct incorporation of semiconducting memory elements into superconductor circuits \cite{mem2} or on the development of Josephson junctions with ferromagnetic barriers -- see, e.g., \cite{spin4, spin1,spin2,spin3,spin5}. While promising in several respects, hybrid structures face many problems related to conversion between different forms of information representation and fabrication difficulties, and still did not reach the level of completely satisfactory practical circuits. The goal of this work is to suggest and demonstrate the main operating principles of all-superconducting memory cells which can be scaled down in size into the range of few tens of nanometers, and which do not suffer from the aforementioned problems. The memory is based on the loops made of superconducting nanowires \cite{Hopkins-Science}, in which information is encoded in the different vorticity states supported by the nanowire loops as long as the loop size exceeds several coherence lengths of the superconductor.

Transport properties of superconducting nanowires are defined by the combination of large kinetic inductance originating from the kinetic energy of the supercurrent flow, and Little's phase slips: abrupt jumps of the phase difference on the wire by $\pm 2\pi$ which are induced by the fluctuation-driven local suppression of the superconducting order parameter -- see, e.g., \cite{rev1, Bezryadin-jpcm,Bezryadin-book} for review. These two factors produce effects that qualitatively mimic the basic characteristic features of the typical SFQ circuits with larger dimensions, e.g., they make the switching dynamics of the nanoloops similar to those of traditional superconducting quantum interference devices (SQUIDs) based on Josephson junctions.  Recently, the main attention in studies of nanowire transport properties has been focused on the quantum \cite{leggett} and Coulomb blockade effects \cite{averin}, which includes the quantum decay of the supercurrent \cite{mqt1,mqt2,mqt3,mqt4}, macroscopic quantum coherence of the magnetic flux \cite{mqc}, and transport of individual Cooper pairs \cite{nazarov2011,zorin,bloch,nazarov2013}. Besides fundamental novelty, the interest attracted by these effects is motivated also by potential applications, in particular to quantum computing \cite{mooij2005} and metrology \cite{mooij2006}, where nanowires hold promise of improved quantum standard of electric current with potentially larger output currents than possible with systems based on tunnel junctions \cite{rmp}.

While applications of superconducting nanowires to quantum computing still need to overcome the uncertainty related to the role and magnitude of dissipation in the dynamics of quantum phase slips, the classical switching dynamics of the nanowire loops studied in this work offer significant, and up to now not fully explored, advantages for classical superconductor-based computing. As explicitly demonstrated below, the nanowire SQUIDs encoding information in the vorticity states have characteristics that make them promising memory elements. We show that the memory (vorticity) states can be manipulated in a controllable fashion. The memory state is written by applying an oscillating current of an appropriate magnitude, at a specified magnetic field. The current and the field are chosen such that only one unique vorticity value remains stable at the peak of the current. The vorticity of the loop remains stable over macroscopically-long time scales, thus offering a nonvolatile nanoscale memory. The state of the memory, i.e., the vorticity of the loop, is read by measuring the critical current. The critical current is vorticity-dependent because the vorticity quantum number defines the value of the persistent supercurrent in the loop. The vorticity can also be measured using microwave techniques \cite{mqt4}, since the winding number has an impact on the kinetic inductance of the device. We also suggest a quantitative model of the energy properties of these SQUIDs which support their potential for operation as dissipation-free memory cells.

Specifically, we performed experiments on two nanowire SQUIDs made of Mo$_{75}$Ge$_{25}$ superconductor. The alloy is amorphous, so the nanowires are free of grain boundaries and thus homogeneous. Device 7715s1 was made using molecular templating. This procedure has been described previously \cite{Murphy,Bezryadin-jpcm}. The wires and electrodes of Device 82915s2 were patterned by electron-beam lithography. To create narrow features, we implemented the technique described in reference \cite{Zailer-Supercond}. High-dosage exposure is used to crosslink the PMMA over the desired pattern of Mo$_{75}$Ge$_{25}$. Next, photolithography is used to create contact pads connected to the electrodes, and the unprotected Mo$_{75}$Ge$_{25}$ is removed by etching with CF$_4$. Both devices consist of two wires (the wires are geometrically different) which connect in parallel to two electrodes, forming a superconducting loop. Device 7715s1 (Fig.\ 1a) has one wire that is 42 nm wide and 140 nm in length, and one wire 26 nm wide and 158 nm in length. The wires are separated by 2.5 $\mu$m. For Device 82915s2 (Fig. 1b), the width of the strip leading to the loop is 210 nm, and the hole in the middle of the loop is 105 nm in width and 145 nm in length. In patterning Device 82915s2, an asymmetry was created by varying the distance between the electrodes. The wires were patterned to be 200 nm and 150 nm in length by programming the electron beam to make a single pass along a straight line for each wire. The dose on the straight line pass was 90~nC~cm$^{-1}$. However, due to the electron beam lithography proximity effect, the wires range in widths from 30 nm to 40 nm and the lengths are shorter than the nominal lengths. Device 82915s2 was purposefully fabricated to be asymmetric so that it can operated as a memory at a constant value of magnetic field, as will be shown. Note that the coherence length $\xi$ of Mo$_{75}$Ge$_{25}$ is known to be on the order of 10 nm although some estimates show that it might be as large as 20 nm in nanowires \cite{Rogachev} due to the expected width dependence. 

The experimental setup has been described in detail previously \cite{Bezryadin-jpcm,Murphy}. Measurements are performed in a He-3 setup with pi-filters installed at the top of the cryostat and copper-powder filters and silver paste microwave radiation filters, both at the base temperature of 300 mK. A current is applied through the device (also at 300 mK) and the voltage difference between the electrodes to which the wires are linked is measured.

Both tested devices show well defined critical currents $I_C$ at temperatures sufficiently below their critical temperatures. The critical currents measured at 300 mK are given in Table I. As the bias current is increased from zero, the voltage across the device is initially zero, but it suddenly jumps from zero to a large non-zero value, indicating that the device has become normal. The current at which this transition takes place is recorded as the switching current I$_{SW}$ (Fig.\ 2). Here we neglect the small differences which might exist between the switching current and the true critical current and assume $I_C = I_{SW}$. In these devices the voltage-current (V-I) curves show a hysteresis because the critical current is significantly larger than the retrapping current, i.e., the current at which the device goes back to the superconducting state as the bias is reduced.

In Figure 2a we plot the switching current of Device 7715s1 as a function of magnetic field. Measurements of the switching current were taken at a sweep frequency of 1.1 Hz. The negative switching currents in this figure are inverted. Black circles show the positive switching current values I$_{SW}$(+), while red points are the magnitude of the negative switching current I$_{SW}$(-). 

Many characteristics of the system can be understood assuming that the current-phase relation is linear. This is a reasonable assumption at temperatures much lower than T$_C$ \cite{Tinkham}. Fitting parameters with this assumption are the critical current I$_{Cj}$ of each wire, $j=1,2$, i.e. the maximum supercurrent that nanowire can support, and the critical phases $\phi_{Cj}$ across the wires that correspond to these currents. They are presented in Table 1, while the fits are shown as solid lines in Figure 2.  We use this one-dimensional model to describe our nanowires because we see no evidence of the hysteresis that would be caused by a vortex getting trapped inside one of the wires. This is consistent with the theoretical expectation that vortices cannot be stable in a thin wire if its width is less than $4.4\xi$ \cite{likharev}. But even if a wire is somewhat wider than this limit, stabilizing a vortex in it would require a field much larger than we ever apply, due to the small area of the nanowire.

Although vortices cannot sit on the nanowires, they can move across a nanowire and enter the loop. Such events constitute phase-slips. The total number of vortices which enter the loop minus the total number which exit the loop represents the main quantum number of the loop, i.e., its vorticity $n_v$ or the winding number of the order parameter. In other words, the vorticity is the number of vortices which are trapped in the nanowire SQUID loop. This number also equals the phase accumulated around the loop divided by 2$\pi$.

Many vorticity states of the nanowire SQUID are metastable at the same applied magnetic field. This is due to the fact that the nanowires achieve critical phase differences considerably larger than $2\pi$ -- see Table 1. We therefore introduce the optimal vorticity $n_{opt}$ of the system defined as the vorticity state with the largest critical current, denoted $I_C(n_{opt})$ at a given value of magnetic field. For the purposes of this work, it is important to reiterate two facts about the optimal vorticity state. First, the optimal vorticity for positive currents and negative currents is not necessarily equal. Second, in the unique vorticity diamond (UVD), (at the largest currents at which the sample is still superconducting \cite{Murphy}), only the optimal vorticity state is stable. The unique vorticity diamond of state $n_v$ = 0 at positive currents is shown as a gray shaded region for both devices in Figure 2.

We demonstrate operation of our devices as memory elements using the vorticity as the information-carrying quantum number. To write a state, we repeatedly drive the system between the UVD of the desired vorticity state and zero current, ensuring that the system never leaves the region in which the desired vorticity is stable (i.e., it should stay within the corresponding Little-Parks diamond) one of which is shown by blue lines in Fig.~\ref{Fig-operation}b \cite{Murphy}.

To read the state, we apply a pulse of current with a maximum greater than the maximum switching current among all possible $n_v$ values, and measure the current at which the system switches to the normal state. The vorticity associated with the switching current branch on which the switching event takes place tells us the vorticity state of the system and therefore the memory state. We determine the vorticity associated with each branch based on the nanowire loop model of Ref.~\cite{Murphy}. We read the state at a magnetic field where the switching currents corresponding to the two possible written states are the optimal and next largest switching currents. We select a field where these two switching currents are similar but distinct. This increases the likelihood that a phase-slip on the non-optimal vorticity branch will be detected as a switching to the normal state. A phase-slip is an event by which the vorticity of the loop changes by one \cite{Murphy,Bezryadin-jpcm}. The switching happens because a single phase slip releases enough heat to suppress the critical current below the applied current.

To operate Device 7715s1, a sinusoidal ac current is applied at one of two magnetic fields to write either state n$_v$ = -1 or n$_v$ = 0. In Fig. 2a the triangle and circle indicate the magnetic fields and the magnitudes of the sinusoidal current used to write the two states. State n$_v$ = -1 is written at the lower field (triangle) of -0.32 G, and state n$_v$ = 0 is written at the higher field (circle) of -0.12 G. In both cases the amplitude of the sinusoidal bias current is set to 43.5 $\mu$A. At -0.12 G, which is the field applied to write state n$_v$ = 0, state n$_v$ = 0 is stable at both -43.5 $\mu$A and +43.5 $\mu$A current. However, while state n$_v$ = -1 is stable at -43.5 $\mu$A, it is not stable at +43.5 $\mu$A current. Such asymmetry with respect to the polarity of the bias current originates from the asymmetry of the nanowire SQUID, namely from the fact that the nanowires have a somewhat different critical phase. In fact, state n$_v$ = 0 is the only vorticity state which is stable at all currents  between -43.5 $\mu$A and +43.5 $\mu$A at -0.12 G. Therefore, over many current oscillations, the system is expected to enter and remain in state n$_v$ = 0. Similarly, at -0.32 G, the system has to enter and remain in state n$_v$ = -1. The current through the device oscillates with a frequency of 101 Hz as the state is written, and writing is performed over two seconds, resulting of approximately 200 writing cycles. (We envision that high frequency microwave pulse will be used in the future generations of memory devices.) To read the state, current is swept from zero to 50 $\mu$A at the read-field B = -0.24 G and the switching current is measured.  A switching event on the maximum switching current branch is interpreted as n$_v$ = 0 state reading; a switching event at the second highest switching current branch was read as state n$_v$ = -1. This reading process appears 100\% reliable, thus the switching always happens at a lower critical current if the vorticity is not the optimal one. In other words, the vorticity cannot adjust itself (without causing a switching to the normal state) to allow a higher critical current in our readout algorithm.

The resulting distributions are shown in Figure 3a. Two distinct distributions of switching currents are found to depend on the written vorticity state. As expected, the higher switching current events are observed in measurements after the system has been written in state n$_v$ = 0 and the lower current events are detected in measurements after the system has been written in state n$_v$ = -1. This experiment achieved \textit{perfect} fidelity, and demonstrates the potential of an asymmetric nanowire SQUID as a memory device.

We operate Device 82915s2 at a constant magnetic field of 424 G (Fig.\ 2b).  To write state n$_v$ = 0, the current is alternated between 0 and 29 $\mu$A fifteen times. To write state n$_v$ = 1, the current is alternated between 0 and -28 $\mu$A fifteen times. The ends of these paths are marked by  triangles and circles in Fig. 2b. Note that the circle at positive currents specifies a point that lies within the UVD of state $n_v$ = 0 at positive currents, i.e. $I_C(n_v = 1) < I_{write} < I_C(n_v = 0)$ where $I_{write}$ is the largest value of the current applied to write state $n_v$ = 0. Similarly, the largest magnitude current applied to write state $n_v$ = 1 lies within the UVD of state $n_v$ = 1 at negative currents. The asymmetry of the device allows us to find a magnetic field where the positive-bias-current and negative-bias-current optimal vorticity states differ by one. The state is read by applying a pulse of positive current from zero to 35 $\mu$A. At B = 424 G, the positive critical currents for states n$_v$ = 0 and n$_v$ = 1 are distinguishable, about 1 $\mu$A apart. This procedure achieved fidelities of 89 \% for state n$_v$ = 1 and 98 \% for state n$_v$ = 0 (Fig.\ 3b) which could be further improved by an optimization of the writing point position.  Note that the noticeable difference in the periods of Device 82915s2 and 7715s1 is due to the large difference in loop size as well as the large difference in the electrode width \cite{Hopkins-Science}. 

We then test the stability of the superconducting state in Device 7715s1 over time by adding a delay between writing the state and reading the state. After writing state n$_v$ = 0 or n$_v$ = -1, the current is reduced to zero and the field is set to its read-value. Then the system waits a given amount of time before being read. We do not find any single unintended flip of the written state (Fig.\ 4). The total duration of the test was more than 2 hours. Thus the rate of phase-slips, both thermal and quantum, in these nanowires at zero bias is less than 1.4 x 10$^{-4}$ s$^{-1}$. Probably it is even many orders lower since we see from previous works that the rate of phase slips is exponentially low at low bias current. This result is in agreement with the previous finding that quantum phase slips do not occur in Mo$_{75}$Ge$_{25}$ nanowires if the bias current is much lower than the critical current and if the wire normal resistance is less than the resistance quantum (6.5 k$\Omega$) \cite{Chu}.

A good memory element needs to satisfy several conditions. It has to be possible to set the device in a desired state and read out this state, with the number of such accessible and distinguishable states being at least two. The devices have to be small, and the reading and writing of the states has to occur on a short time scale, while the states have to be stable over macroscopically-long time, ideally without external power applied. It has to be possible to integrate the devices into large-scale circuits, in which they do not interfere with each other. Scaling of these circuits to a larger number of elements would be simplified considerably if the power dissipation during the reading and writing steps is negligible. In this work, we have addressed many of these requirements. We have shown that we can read and write two distinct memory states and that the memory is stable over long periods of time. Our studies of Device 82915s2 also show that memory functions can be performed on a sub-micron scale device. The small geometric inductance of Device 82915s2 (which we estimate to be on the order of 0.1 - 1 pH) is insufficient to produce metastable vorticity states \cite{Tinkham}. However, since the states of the nanowire loop, including the multiple metastable vorticity values, are defined mostly by the kinetic inductance of the nanowires, which in our devices is on the order of 100 pH, the dimensions of the loop can be reduced \cite{McCaughan}, in contrast to memory devices based on common SQUIDs composed of Josephson junctions. \cite{spin2}. Since the kinetic inductance increases with decreasing cross-sectional dimensions of the wire, nanowire SQUIDs as memory elements can still be reduced further into the size range of tens of nanometers. The aspects of the nanoloop memory not addressed in this work, measurements of the switching time, and study of the larger arrays of the nanowire SQUIDs can be addressed in further work, and should not create obstacles for development of the nanoscale superconducting memory elements. Due to the simplicity in our electron-beam lithography pattern, we believe that superconducting memories composed of nanowire loops should be readily scalable for dense operation, although we have not studied yield or scalability directly. To operate at 4 K, we suggest using a material with high enough critical temperatures $T_C$ such that the critical currents of non-optimal vorticity states are measurable at this temperature. Such materials might be, for example, NbTi ($T_C$ = 9 K), NbTiN ($T_C$ = 13 K), or MoSi ($T_C$ = 7 K), which we plan to test in the future. The required high speed of operations should be possible since the attempt frequency is $\sim$~100~GHz in such nanowires.

As the last section, we discuss the power dissipated in the reading and writing steps of the nanoloop memory by considering the energy of the nanowire loop. In the linear approximation used to describe the current-phase relation of the loop wires, the energy $U$ of the nanoloop is the sum of the internal "kinetic" energy of the loop, and the energy of interaction with the bias current $I$. Using this fact, one can derive the following expression for $U$: 
\begin{equation}
U = \frac{(\Phi_0/2\pi)^2}{2}[\frac{\phi_1^2}{l_1}+\frac{\phi_2^2}{l_2}] - \frac{\Phi_0 I}{l_1+l_2}[n_1 l_2 + n_2 l_1] \, ,
\label{en} \end{equation}
where $\Phi_0$ is the magnetic flux quantum, $l_j$ is the kinetic inductance of nanowire $j$, $\phi_j$ is the phase difference of the superconducting order parameter between the two ends of the wire, and $n_j$ is the total number of the phase slip tunneling events which have occured in it. An interesting aspect of this expression is that it represents the exact magnetic dual of the electrostatic Coulomb energy of the
single-electron transistor \cite{al}. This implies, for example, that the phase-slip tunneling events can be manipulated in the same way as the tunneling of individual electrons in the single-electron structures. 

The inductances $l_j$ in the energy $U$ given by Eq.~\ref{en} are related to the critical currents by $I_{Cj} = \phi_{Cj} \Phi_0/(2\pi l_j)$, and from the values in Table 1, are estimated to be on the order of 100 pH. Since the geometric inductances of our wires are at least two orders of magnitude smaller than this, we see that the loop inductance is dominated by the kinetic contribution from the supercurrent flow. The distribution of the bias current into the two wires, $I = I_1 + I_2$ is determined by these inductances, $I_j=Il_{j'}/(l_1+l_2)$, where $j'\neq j$. The magnetic field $B$ creates the difference $\phi_B$ between the phases $\phi_j$ in addition to the difference associated with the vorticity state $n=n_2 - n_1$ of the loop, $\phi_1 - \phi_2 = 2 \pi n - \phi_B$. Note that the Meissner phase bias $\phi_B$ is related to the applied field by $\phi_B = 2 \pi B / \Delta B$ where $B$ is the magnetic field and $\Delta B$ is the period \cite{Hopkins-Science}. Using these relations to determine the phases $\phi_j$, we can find the change in energy (\ref{en})
in a phase-slip tunneling event in the first, $n_1\rightarrow n_1\pm 1$, or the second, $n_2\rightarrow n_2\pm 1$, wire. The energy change vanishes when the loop bias satisfies the conditions
\begin{equation}
\phi_B^{(1)} = \frac{2 \pi I l_2}{\Phi_0} + 2 \pi (n \mp \frac{1}{2})\, , \;\;\;\;\; 
\phi_B^{(2)} = -\frac{2 \pi I l_1}{\Phi_0} + 2 \pi (n\pm \frac{1}{2})\, , 
\end{equation}
in the first and the second wire, respectively. If the phase-slip transitions are manipulated  
in the vicinity of the bias conditions satisfying these relations, similarly to what can be done with individual electrons \cite{rev}, the phase slips will not dissipate energy. In this way, one can create low-dissipation memory cells out of the nanowire SQUIDs.

In summary, we have demonstrated that small superconducting nanowire loops (much smaller than 1 $\mu$m) can be used as superconducting memory elements, which reliably store information without any power supplied. Information is encoded in the vorticity of the system, which is multivalued due to the significant kinetic inductance of the nanowires. We were able to read and write the states using current pulses and magnetic field. At this time memory read and write operations take time on a scale of a second. In the future we plan to operate the memory using microwave pulses to bring the time scale to a fraction of a nanosecond. Note that microwave readout of the vorticity of nanowire loops has been shown in Ref. \cite{mqt4} We have also proposed a strategy of reducing energy dissipation in memory operation to a near-zero value. 

We thank M. Manheimer and J. Ku for helpful discussions. This work was supported by the National Science Foundation under the Grant No. ECCS-1408558.

\pagebreak

\begin{table}[]
\centering
\label{Table}
\begin{tabular}{ | l | l | l | l | l | l | l | l |}
\hline
Device  & $R_N$ ($\Omega$)& $T_{C, wires}$  (K)& $T_{C, film}$  (K)&I$_{C1}$ ($\mu$A) & I$_{C2}$ ($\mu$A) & $\phi_{C1}$ (rad) & $\phi_{C2}$ (rad) \\
\hline

7715s1  & 320 & 5.5 & 6.1 & 16.9              & 31.1              & 23.6        & 21.1        \\
82915s2 & 230 & 5.0 & 5.5 & 12.3              & 19.4              & 13.7        & 15.9     \\
\hline
\end{tabular}
\caption{Device Properties and Fitting Parameters. $R_N$ is the normal resistance of the nanowire loop. $T_{C, wires}$ is the superconducting transition temperature of the nanowires, defined empirically as the midpoint of the transition.  $T_{C, film}$ is the superconducting transition temperature of the electrodes and contact pads. I$_{C1}$ and I$_{C2}$ are the critical currents of the two nanowires forming the superconducting loop, and $\phi_{C1}$ and $\phi_{C2}$ are the critical phases of the nanowires forming the loop. The critical currents and phases are found by performing best fits to the data.}
\end{table}

\begin{figure}[t!]
  \includegraphics[width=15cm]{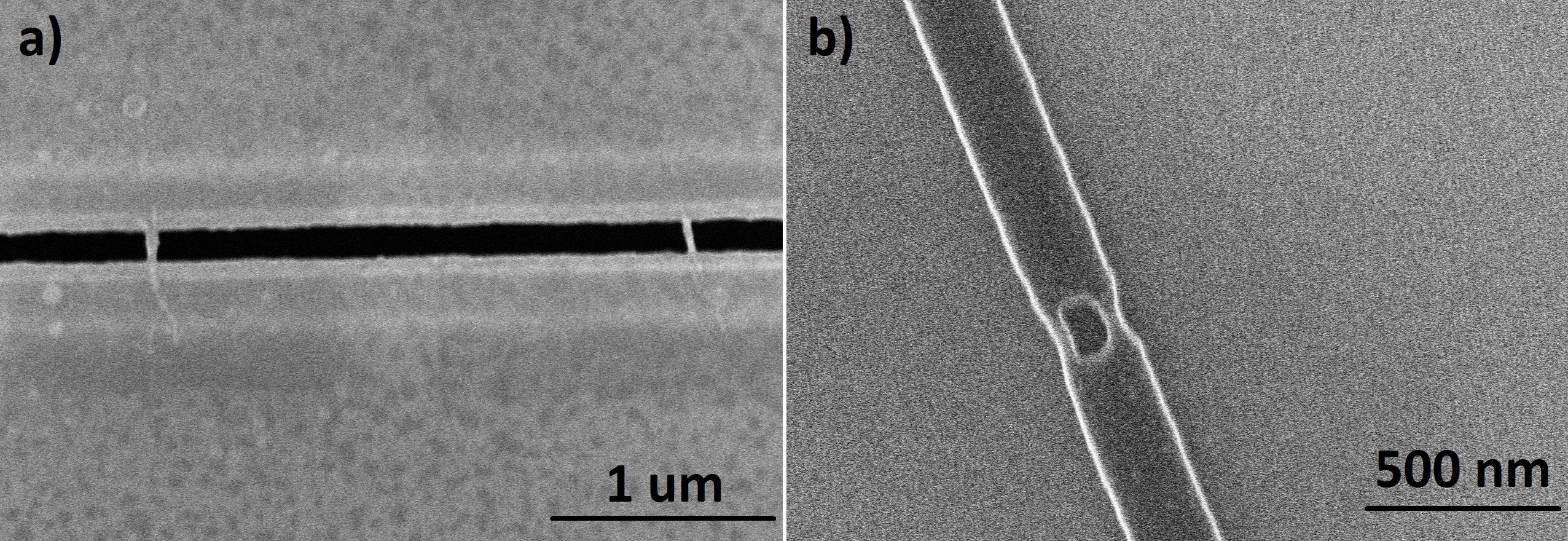}
  \caption{ a) An SEM image of Device 7715s1. Two carbon nanotube templated Mo$_{75}$Ge$_{25}$ wires lay across a roughly 150 nm wide trench, 2.5 $\mu$m apart. The two wires have similar dimensions, but are not identical. b) An SEM image of Device 82915s2. The Mo$_{75}$Ge$_{25}$ (dark) is patterned into two geometrically different nanowires sitting 150 nm apart. The right wire is made shorter.
 }\label{Fig-sem}\vskip-.5cm
\end{figure}

\begin{figure}[t!]
\includegraphics[width=8cm]{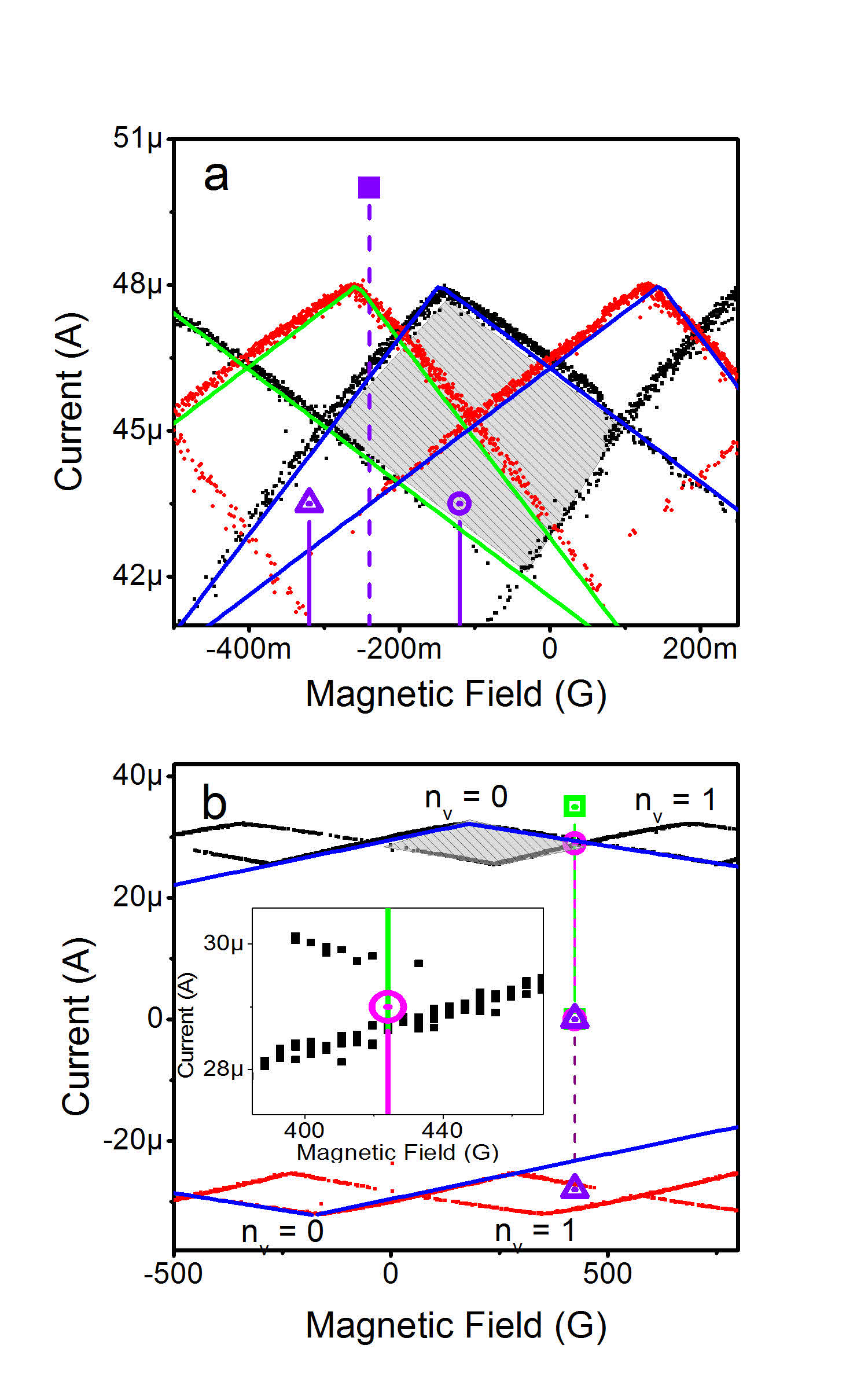}
\caption{Memory Operation. a) Device 7715s1. Black and red dots represent, respectively, positive  I$_{SW}$(+) and negative I$_{SW}$(-) switching currents, the latter inverted in the plot. Theoretical fits \cite{Murphy} to the boundaries of vorticity states n$_v$ = 0 and n$_v$ = -1 are shown by lines. To write state n$_v$ = -1, we apply a sinusoidal current sweep with an amplitude of 43.5 $\mu$A and at a field B = -0.32 G (triangle). To write state n$_v$ = 0, we apply a sinusoidal ac current sweep with an amplitude of 43.5 $\mu$A  and at a field B = -0.12 G (open circle). To read the state we apply a positive current sweep at B = -0.24 G (purple square) exceeding the maximum switching current. The state of the system is determined by the branch at which the system switches to the normal state. b) Device 82915s2. Black dots represent I$_{SW}$(+) and red dots represent I$_{SW}$(-). The blue curve shows a theoretical fit \cite{Murphy} of the critical current to state $n_v$ = 0. Writing and reading are performed at a constant field of 424 G. The vorticities of the UVD for states $n_v=0$ and $n_v=1$ are labeled on the plot. We either write state $n_v$ = 0 by oscillating the current between zero and the UVD for state $n_v$ = 0 at positive currents (pink circle), or we write state $n_v$ = 1 by oscillating the current between zero and the UVD for state $n_v$ = 1 at negative currents (purple triangle). Reading is performed by sweeping the current from zero to 35 $\mu$A (green square). In the inset we zoom in on the current and field at which state $n_v=0$ is written.
}\label{Fig-operation}
\end{figure}

\begin{figure}[t!]
  \includegraphics[width=16cm]{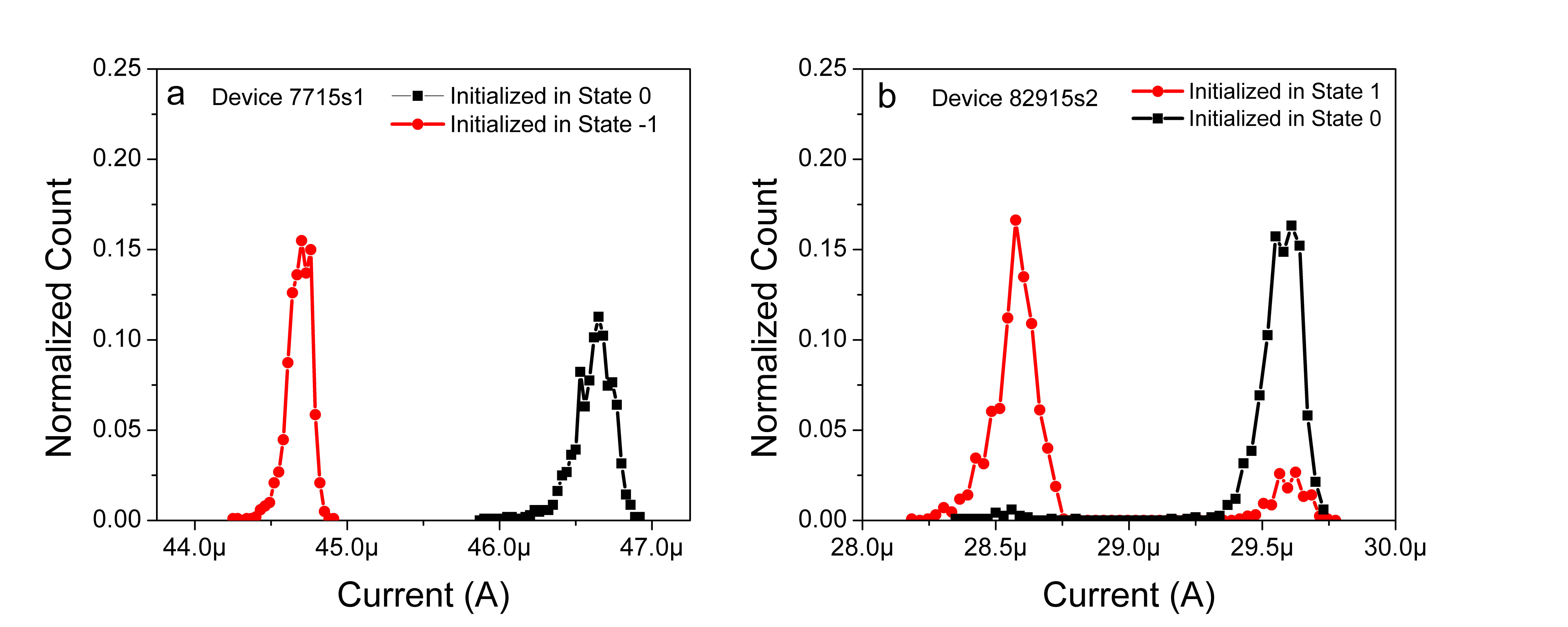}
  \caption{Readout Distributions. a) The readout distributions measured after writing Device 7715s1 in state n$_v$ = 0 (black) and state n$_v$ = -1 (red) exhibit perfect fidelity after 1,000 measurements. b) The readout distributions read after writing Device 82915s2 in state n$_v$ = 0 (black) and state n$_v$ = 1 (red) exhibit fidelities of 98 \% and 89 \% respectively after 1,000 measurements. Each distribution is binned and normalized by the total number of events.
 }\label{Fig-dists}\vskip-.5cm
\end{figure}

\begin{figure}[t!]
  \includegraphics[width=15cm]{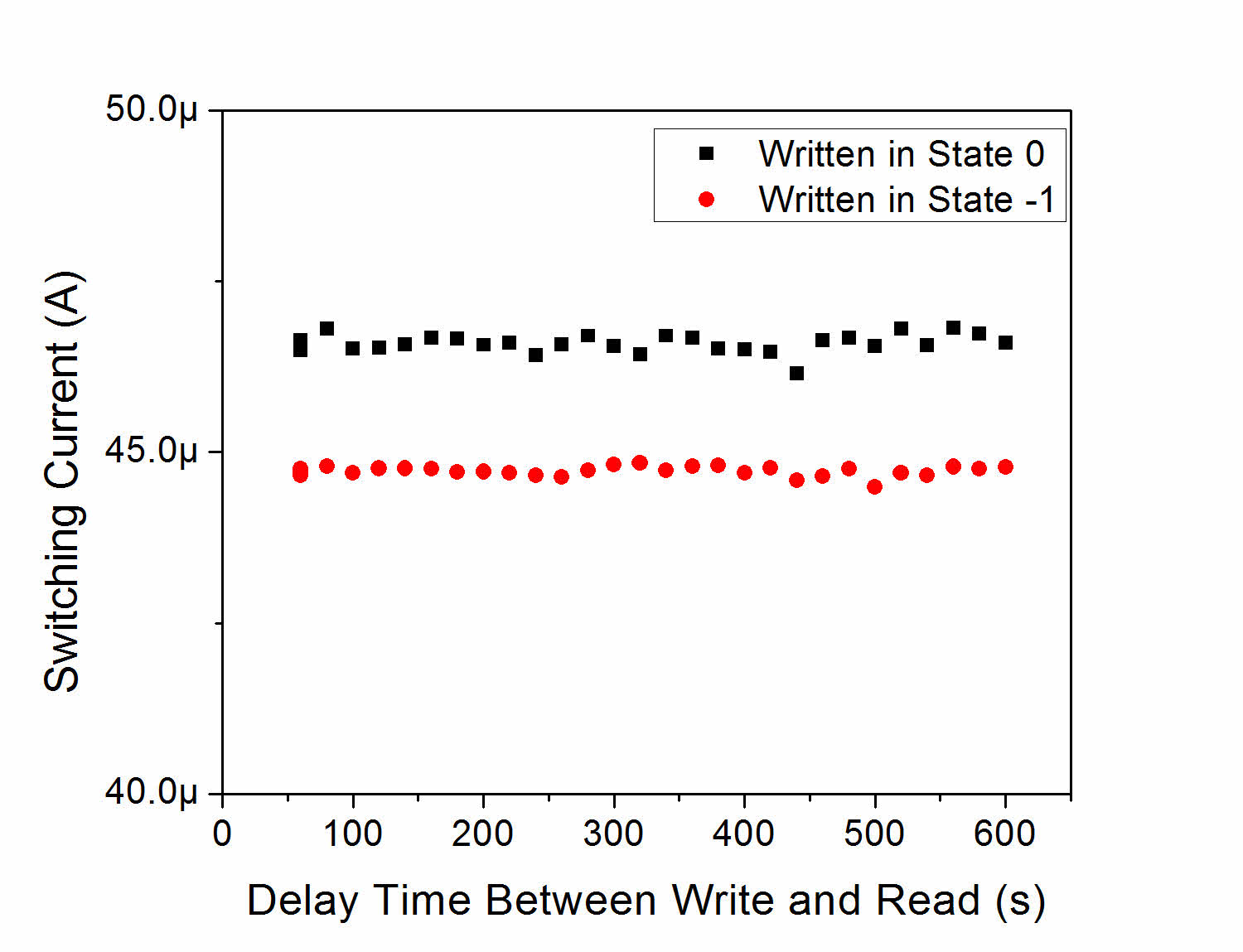}
  \caption{Memory stability of Device 7715s1. A delay is added between writing and reading the state. Both states n$_v$ = 0 and n$_v$ = -1 retain their vorticity up to ten minutes after they are written, for many cycles. The total test time was more than 2 hours and no single memory-state unintended flipping occurred. This shows also that there is no noise level in the setup which would cause unintended phase slips.
 }\label{Fig-delay}\vskip-.5cm
\end{figure}


\begin{thebibliography}{00}

\bibitem{spin4} B. Baek, W.H. Rippard, S.P. Benz, S.E. Russek, P.D. Dresselhaus, \textit{Nat. Comm.} \textbf{5}, 3888 (2014).

\bibitem{Aprili} M.~Aprili, \textit{Nat. Nanotech.} \textbf{1}, 15 (2006).

\bibitem{Hopkins-Science} D. S. Hopkins, D. Pekker, P. M. Goldbart, A. Bezryadin, \textit{Science} \textbf{308}, 1762 (2005).

\bibitem{Manheimer} M. Manheimer, Cryogenic Computing Complexity (C3), IARPA-BAA-13-05, https://www.iarpa.gov/index.php/research-programs/c3

\bibitem{rsfq} K. K. Likharev, V. K. Semenov, \textit{IEEE Trans. Appl. Supercond.} \textbf{1}, 3 (1991).

\bibitem{Filippov} T. V. Filippov \textit{et al.}, \textit{Physics Procedia} \textbf{36}, 59 (2012).

\bibitem{nsquid} J. Ren, V. K. Semenov, Yu. A. Polyakov, D. V. Averin, J. S. Tsai, \textit{IEEE Trans. Appl. Supecond.} \textbf{19}, 961 (2009).

\bibitem{rql} Q. P. Herr, A. Y. Herr, O. T. Oberg, A. G. Ioannidis, \textit{J. Appl. Phys.} \textbf{109}, 103903 (2011).

\bibitem{esfq} M. H. Volkmann, I. V. Vernik, O. A. Mukhanov, \textit{IEEE Trans. Appl. Supercond.} \textbf{25}, 1301005 (2015).

\bibitem{mem0} S. V. Polonsky, A. F. Kirichenko, V. K. Semenov, K. K. Likharev, \textit{IEEE Trans. Appl. Supercond.} \textbf{5}, 3000 (1995).

\bibitem{mem1} S. Nagasawa, H. Numata, Y. Hashimoto, S. Tahara, \textit{IEEE Trans. Appl. Supercond.} \textbf{9}, 3708 (1999).

\bibitem{mem2} T. van Duzer \textit{et al.}, \textit{IEEE Trans. Appl. Supercond.} \textbf{23}, 1700504 (2013).

\bibitem{spin1} A. K. Feofanov \textit{et al.}, \textit{Nat.\ Phys.} \textbf{6}, 593 (2010).

\bibitem{spin2} E. Goldobin \textit{et al.}, \textit{Appl. Phys. Lett.} \textbf{102}, 242602 (2013).

\bibitem{spin3} I. V. Vernik \textit{et al.}, \textit{IEEE Trans. Appl. Supercond.} \textbf{23}, 1701208 (2013).

\bibitem{spin5} B. M. Niedzielski, E. C. Gingrich, R. Loloee, W. P. Pratt, N.O. Birge, \textit{Supercond. Sci. Technol.} \textbf{28}, 085012 (2015).

\bibitem{rev1} K. Yu. Arutyunov, D. S. Golubev, A. D. Zaikin, \textit{Phys. Rep.} \textbf{464}, 1 (2008).

\bibitem{Bezryadin-jpcm} A. Bezryadin, \textit{J. Phys.: Condens. Matter} \textbf{20}, 043202 (2008).

\bibitem{Bezryadin-book} A. Bezryadin, "Superconductivity in Nanowires: Fabrication and Quantum Transport", WILEY-VCH, (2013).

\bibitem{leggett} A. O. Caldeira, A. J. Leggett, \textit{Phys. Rev. Lett.} \textbf{46}, 211 (1981).

\bibitem{averin} D. V. Averin, A. B. Zorin, K. K. Likharev, \textit{Sov. Phys. JETP} \textbf{61}, 407 (1985).

\bibitem{mqt1} N. Giordano, \textit{Phys. Rev. Lett.} \textbf{61}, 2137 (1988).

\bibitem{mqt2} A. Bezryadin, C. N. Lau, M. Tinkham, \textit{Nature} \textbf{404}, 971 (2000).

\bibitem{mqt3} M. Sahu \textit{et al.}, \textit{Nat. Phys.}  \textbf{5}, 503 (2009).

\bibitem{mqt4} A. Belkin, M. Belkin, V. Vakaryuk, S. Khlebnikov, A. Bezryadin, \textit{Phys. Rev. X} \textbf{5}, 021023 (2015).

\bibitem{mqc} O .V. Astafiev \textit{et al.}, \textit{Nature} \textbf{484}, 355 (2012).

\bibitem{nazarov2011} A. M. Hriscu, Yu. V Nazarov, \textit{Phys. Rev. B}  \textbf{83}, 174511 (2011).

\bibitem{zorin}  T. T. Hongisto, A. B. Zorin, \textit{Phys. Rev. Lett.} \textbf{108},  097001 (2012).

\bibitem{bloch} J. S. Lehtinen, K. Zakharov, K. Yu. Arutyunov, \textit{Phys. Rev. Lett.} \textbf{108}, 187001 (2012).

\bibitem{nazarov2013} A. M. Hriscu, Yu.V Nazarov, \textit{Phys. Rev. Lett.}  \textbf{110}, 097002 (2013).

\bibitem{mooij2005} J. E. Mooij, C. J.P .M. Harmans, \textit{New J. Phys.} \textbf{7}, 219 (2005).

\bibitem{mooij2006} J. E. Mooij, Y. V. Nazarov, \textit{Nat. Phys}. \textbf{2}, 169 (2006).

\bibitem{rmp} J. P. Pekola \textit{et al.}, \textit{Rev. Mod. Phys.} \textbf{85}, 1421 (2013).

\bibitem{Murphy} A. Murphy, A. Bezryadin, arXiv:1609.03877 [cond-mat.supr-con] (2016).

\bibitem{Zailer-Supercond} I. Zailer, J. E. F. Frost, V. Chabasseur-Molyneux, C. J. B. Ford, M. Pepper, \textit{Supercond. Sci. Technol.} \textbf{11}, 1235 (1996).

\bibitem{Rogachev} A. Rogachev, T.-C. Wei, D. Pekker, A. T. Bollinger, P. M. Goldbart and A. Bezryadin, \textit{Phys. Rev. Lett.} \textbf{97}, 137001 (2006).

\bibitem{Tinkham} M. Tinkham, \textit{Introduction to Superconductivity}, 2nd ed. (McGraw-Hill,
New York, 1996).

\bibitem{likharev} K.~K.~Likharev, Rev. Mod. Phys. \textbf{51}, 101 (1979).

\bibitem{Chu} S. L. Chu, A. T. Bollinger, A. Bezryadin, \textit{Phys. Rev. B} \textbf{70}, 214506 (2004).

\bibitem{McCaughan} A.N. McCaughan, Q. Zhao, K.K. Berggren, \textit{Sci. Rep.} \textbf{6}, 28095 (2016).

\bibitem{al} D.V. Averin and K.K. Likharev, J.\ Low Temp.\ Phys.  \textbf{62}, 345 (1986).

\bibitem{rev} D.V. Averin and J.P. Pekola, Europhys.\ Lett. \textbf{96}, 67004 (2011).


\end{thebibliography}
\end{document}